\documentclass[twocolumn,showpacs,preprintnumbers,amsmath,amssymb,epsf,psfig]{revtex4}
\usepackage{graphicx}
\usepackage{dcolumn}
\usepackage{bm}

\begin{document}

\title{An Optical Analog of a Black Hole}
\author{Andrew Royston and Richard Gass}
\affiliation{Department of Physics, University of Cincinnati,
Cincinnati,  OH 45220-0011} 

\date{\today}

\begin{abstract}               
Using media with extremely low group velocities one can create an optical analog of a curved space-time. Leonhardt and Piwnicki have proposed that a vortex flow will act as an optical black hole. We show that although the  Leonhardt - Piwnicki flow has an orbit of no return and an infinite red-shift surface, it is not a true black hole since it lacks a null hypersurface.  However a radial flow will produce a true optical black hole that has a Hawking temperature and obeys  the first law of black hole mechanics. By combining the  Leonhardt - Piwnicki  flow with a radial flow we obtain the analog of the Kerr black hole.

\end{abstract}
\maketitle

It has long been recognized that a moving dielectric acts as an effective curved space-time for light \cite{Fresnel:1818,Lorentz:1895,Gordon:1923}. This was of only academic interest until recently, when it became possible to make materials with very low group velocities \cite{Hau:1999}. Shortly after this Leonhardt and Piwnicki \cite{LeonhardtPiwnicki:1999,LeonhardtPiwnicki:2000} realized that these materials provided a laboratory test bed for general relativity. It is not of course possible to test the field equations, but one can investigate phenomena which depend solely on the underlying geometry. Black holes are the obvious example both because of their importance, and because black hole physics depends on the geometrical properties of null hypersurfaces, not the field equations.
In a moving dielectric  with a (three-dimensional) fluid flow $\mathbf{u}$, photons follow null geodesics of \begin{equation} ds^{2}=g_{\mu \nu} dx^{\mu} dx^{\nu} \mbox{ where } dx^{\mu}=(c dt, d\mathbf{x}) . \end{equation} The metric $g^{\mu \nu}$ is given by \begin{equation}g^{\mu \nu}= \eta^{\mu \nu}+\left( \begin{array}{cc}\alpha u^{2}/c^{2}&\alpha \mathbf{u}/c \\
 \alpha \mathbf{u}/c&4\alpha \mathbf{u}\otimes \mathbf{u}/c^{2}\end{array} \right),\end{equation} where $\mathbf{u}$ is the fluid flow, $\alpha$ is related to the group velocity $v_{g}$ by $\alpha=c/v_{g}-1$ and $\otimes$ denotes the three-dimensional  direct  product \cite{Gordon:1923,LeonhardtPiwnicki:1999}. This result can be generalized to \begin{equation}g^{\mu \nu}= \tilde{g}^{\mu \nu}+\left( \begin{array}{cc}\alpha u^{2}/c^{2}&\alpha \mathbf{u}/c \\
 \alpha \mathbf{u}/c&4\alpha \mathbf{u}\otimes \mathbf{u}/c^{2}\end{array} \right), \label{metric} \end{equation} where $\tilde{g}^{\mu \nu}$ is the metric of the physical space-time. In principle the underlying physical space-time can be curved, but we will restrict ourselves to flat physical space-times. This does not mean  $\tilde{g}^{\mu \nu}$  is  the standard Minkowski metric because we will be working in curvilinear coordinates. 

Leonhardt and Piwnicki work in cylindrical coordinates and choose a fluid flow given by \begin{equation} \mathbf{u}=\frac{\mathcal{W}}{r} \mathbf{e}_{\varphi} \end{equation} where $ 2\pi  \mathcal{W}$ is the vorticity. This flow yields a metric given by \begin{equation}g^{\mu \nu}=\left( \begin{array}{cccc}1+\frac{\alpha  \mathcal{W}^{2}}{c^{2} r^{2}}&0&\alpha \frac{\mathcal{W}}{c r} &0\\ 
0&-1&0&0 \\
 \frac{ \alpha \mathcal{W}}{c r}&0& \frac{4 \alpha \mathcal{W}^{2}}{c^{2} r^{2}}-1 &0 
\\0&0&0&-\frac{1}{r^{2}}\end{array} \right)
.\end{equation} This metric is slightly different from the one given by Leonhardt and Piwnicki \cite{LeonhardtPiwnicki:2000} since we have explicitly included the factor of $1/r^{2}$ associated with cylindrical coordinates. 
By looking at the turning points of the radial motion Leonhardt and Piwnicki  \cite{LeonhardtPiwnicki:2000} show that there are two critical radii
 $$r_{-}=2 \frac{\mathcal{W}}{c} \left( \frac{c}{v_{g}} \right)^{3/4} \mbox{and } r_{+} =\frac{\mathcal{W}}{c}\left(\frac{c}{v_{g}} \right)^{1/4}. $$  Leonhardt and Piwnicki call these the weak Schwarzschild radius and the Schwarzschild radius respectively.
It is straightforward to compute $g_{00}$ and to show that $r=2 \sqrt{\alpha} \mathcal{W} /c $ is an infinite red shift surface. The infinite red shift surface lies inside the weak Schwarzschild radius for all $\alpha$. For $$\alpha > \left( 1+ \sqrt {65}\right)/32$$ the infinite red shift surface also lies inside the  Schwarzschild radius.

The Leonhardt and Piwnicki  metric does not however, represent a true black hole since it does not possess a  nontrivial null hypersurface and thus will not have a horizon \cite{Hawking:1972}.  A hypersurface $S$ is null if the vector $\mathbf{n}$ which is normal to $S$ is  null.
Since the metric is stationary and axially symmetric we take
\begin{equation} n_{\alpha}=\left( 0,\frac{\partial S}{\partial r},\frac{\partial S}{\partial \varphi} ,0\right) .\end{equation}
 The requirement that $n_{\alpha} n^{\alpha}=0$ gives 
\begin{equation}  \left( \frac{4 \mathcal{W}^{2} \alpha}{c^{2}r^{2}}-1 \right) \left( \frac{\partial S(r,\varphi)}{\partial \varphi} \right)^{2}- \left( \frac{\partial S(r,\varphi)}{\partial r} \right)^{2}=0 . \end{equation} This equation has no real solutions except for the trivial solution $S= \mbox{constant}$.

However, by choosing a radial flow one can get a optical black hole with a horizon.  Working in spherical coordinates we choose a flow \begin{equation} \mathbf{u}= \left( f(r),0,0 \right)  \label{uflow}\end{equation} 
and specialize to the case \begin{equation}f(r)= \beta/\left(r+\epsilon_{0} \right) \label{flow}. \end{equation} Here $\beta$ is a flow parameter, which we will take to be positive, and $\epsilon_{0}$ is a cutoff which prevents the velocity from becoming infinite at the origin. As we will see none of the essential physics depends on $\epsilon_{0}$. There is nothing special about our choice for $f(r)$, there are other choices which will give similar results. With  $f(r)$ given by  Eq. \ref{flow} it is straightforward to show  that the metric has an infinite red shift surface at \begin{equation} r_{H}= \frac{2 \sqrt\alpha\beta}{c}-\epsilon_{0} .\end{equation}  The radial flow metric does have a null hypersurface and thus a true horizon. Since the metric is spherically symmetric we take$$n_{\alpha}=\left( 0,\frac{\partial S}{\partial r},0 ,0\right).$$  The requirement that  $n_{\alpha} n^{\alpha}=0$ then gives
\begin{equation} \left(\frac{4\alpha \beta^{2}}{c^2 \left(r+\epsilon_{0} \right)^{2}}-1 \right)\left( \frac{\partial S}{\partial r}\right)^{2}=0 . \end{equation} which is satisfied when $r=r_{H}$. Thus, for the radial flow metric given by Eqs. \ref{metric}, \ref{uflow} and \ref{flow}, the horizon and the infinite red shift surface coincide.  The flow velocity at the horizon is safely subluminal and is 
$$u_{r_{H}}=\frac{c}{2 \sqrt \alpha}.$$

Since the radial flow optical black hole has a horizon it will have a surface gravity and a Hawking temperature.  The surface gravity $\kappa$ of a black hole is \begin{equation} \kappa^{2} = -\frac{1}{2} \nabla_{\alpha} \ell_{\beta} \nabla^{\alpha} \ell^{\beta} \label{kappa} \end{equation} where $\ell$ is the Killing vector that generates the horizon \cite{DeFelice&Clarke}. Evaluating Eq.\ref{kappa} on the horizon gives the surface gravity on the horizon. For the radial optical black hole the Killing vector that generates the horizon is $\ell=k$ where $$k=\frac{\partial}{\partial t}. $$Evaluating Eq.\ref{kappa} on the horizon gives \begin{equation} \kappa = \frac{c^{3}}{\alpha \beta} \label{surfacegravity} \end{equation}  where we have restored the factors of $c$ that come from the fact that $dx^{0}= c dt .$ Note that $\kappa$ is constant on the horizon as is required by the zeroth law of black hole mechanics. The Hawking temperature of a black hole \cite{Hawking:1975} is \begin{equation} T_{H}=\frac{ \hbar \kappa}{2 \pi k_{B} c}. \end{equation} which gives the optical black hole a Hawking temperature of $$T_{H}=\frac{\hbar c^{2}}{2 \pi \alpha \beta k_{B}}.$$

 Like a normal black hole the optical black hole has a negative specific heat. Larger values of  $\alpha$ will give larger Schwarzschild radii, but cooler Hawking temperatures.

 If we choose $\alpha=2 \times10^{7}$ which corresponds to a physically reasonable group velocity of $v_{g}=15 m/s$ and $\beta=1$, we get $r_{H} \approx 3  \times10^{-5} \mbox{meters} -\epsilon_{0}$, and a Hawking temperature of $T_{H} \approx 5 \mbox{ mK}$. Although this is an enormous  temperature compared to the Hawking temperature of a solar mass black hole ($T_{H} \approx 6 \times 10^{-8} \mbox{K}$), the prospects for experiential observation appear dim at best.

In order to be non-pathological, the metric must describe a space-time that is singularity free in the exterior of the horizon. To check for singularities, we have computed the Kretschamann invariant $$I = R_{\alpha \beta \mu \nu} R^{\alpha \beta \mu \nu}$$  which is a standard method of probing the singularity structure of a space-time \cite{Joshi:1993}. We find $$ I=\frac{4\,{\alpha }^2\,{\beta }^4 F(r)}{c^{4} r^{4}\left(r+\epsilon_{0} \right)^{8}G(r)^{4}} $$ where 
$ F(r) $ is a long polynomial in $r$
and \begin{equation} 
\begin{split}
G(r)&=
 -3\,c^2\,\alpha \,{\beta }^2\,{\left( r + {{\epsilon }_0} \right) }^2 \\ &+ 
        c^4\,{\left( r + {{\epsilon }_0} \right) }^4 + 
        {\alpha }^2\,{\beta }^2\,
         \left( -4\,{\beta }^2 + c^2\,{\left( r + {{\epsilon }_0} \right) }^2 \right) 
      \end{split}
\nonumber
. \end{equation} For positive $\beta$,  $G(r)$ has one positive real root. 
The root is  \begin{equation} 
r_{\mbox{singular}}=\frac{\beta {\sqrt{\alpha }}\,{\sqrt{3 - \alpha  + 
        {\sqrt{25 + \left( -6 + \alpha  \right) \,\alpha }}}}
      }{{\sqrt{2}}\,c} .
\end{equation} However, this singularity always lies inside of the horizon and thus cosmic censorship is preserved.  The singularity is similar to the ring singularity in the Kerr solution.
In addition the space-time has a singularity at $r=0$ in the limit that $\epsilon_{0} \rightarrow 0$

In order to make the connection to the four laws of black hole mechanics \cite{BardeenCarterHawking:1973}, we must calculate the mass associated with the optical black hole. For an asymptotically flat space-time the mass as measured from infinity is
\begin{equation}
M=-\frac{1}{4 \pi G}\int \limits_{\partial S}^{\null} \nabla^{\mu} k^{\nu} \, d\Sigma_{\mu \nu},
\end{equation}
where $d\Sigma_{\mu \nu}$ is the surface element of $\partial S$ and $\partial S$ is the boundary of a space-like hypersurface $S$ and consists of $\partial B$ and $\partial S_{\infty}$. The boundary $\partial B$  is the intersection of $S$ and the horizon, and  $\partial S_{\infty}$ is a 2-surface at infinity. Thus,
\begin{equation} M=-\frac{1}{4 \pi G} \left(\,\int \limits_{\partial B}^{\null} \nabla^{\mu} k^{\nu} \, d\Sigma_{\mu \nu}+
\int \limits_{\partial S_{\infty}}^{\null} \nabla^{\mu} k^{\nu} \, d\Sigma_{\mu \nu}\, \right) \label{mass}. \end{equation}
The first term in Eq. \ref{mass} is the mass associated with the horizon while the second term is the mass outside the horizon as measured at infinity. We can rewrite the first term in Eq. \ref{mass} as \begin{equation} M_{H}= \frac{1}{4\pi G} \, \int \limits_{\partial B}^{\null} \kappa \,da  , \label{holemass} \end{equation} while the second term gives \begin{equation} M_{\mbox{exterior}}= \frac{1}{4\pi G} \, \int \limits_{\partial S_{\infty}}^{\null} \kappa \,da \label{exterior} \end{equation} where $\kappa$ is evaluated on $\partial B$ and $\partial S_{\infty}$ respectively \cite{BardeenCarterHawking:1973} and $da$ is the area element. Evaluating Eq. \ref{holemass} gives \begin{equation} M_{H}= 4 \frac{ \beta c}{G} \end{equation} where we have set $\epsilon_{0}=0$ for simplicity. For $\beta=1$ the effective mass of the black hole is $M_{H} \approx 1.8 \times 10^{19}\, \mbox{Kg} \approx 1 \times 10^{-4} M_{\oplus}$.  The exterior mass, as seen at infinity is zero since for our flow the surface gravity $\kappa$ falls off as $1/r^{6}$. Of course, in any real experimental situation the space will be compact rather than asymptotically flat, and thus there will be an additional contribution to the mass.  Although we will  assume an asymptotically flat space-time, this opens up the possibility of  investigating space-times whose topologies  are not $R^{4}$.

By combining the radial flow with the vortex flow of Leonhardt and Piwnicki \cite{LeonhardtPiwnicki:2000}, we obtain the analog of a Kerr black hole. In spherical coordinates the fluid flow is $\mathbf{u}=\left(f(r),0,\mathcal{W}/r \right)$.  Although the resulting metric is complicated one can show that there is an infinite red shift surface at \begin{equation} r_{H}= \ \frac{2\,\alpha \,{\beta }^2\,{\csc (\theta)}^2}
   {{\sqrt{\alpha \,{\beta }^2\,{\csc (\theta)}^2\,
        \left(c^2\,{\csc (\theta)}^2  -4\,{\mathcal{W} }^2\,\alpha  \right) }}}-\epsilon_{0} .\end{equation} Just as in the Kerr case, the infinite red shift surface is no longer spherical. The infinite redshift surface is well defined only if $|\mathcal{W}| <c/\left(2\sqrt{\alpha} \right)$. For values of $|\mathcal{W}| >c/\left(2\sqrt{\alpha} \right)$, the sign of $g^{\phi \phi}$ changes, which changes the signature of the metric.  When $0<|\mathcal{W}| <c/\left(2\sqrt{\alpha} \right) $ the hole will possess an ergosphere.  The location of the  horizon  does not depend on the vorticity and is unchanged from the purely radial case.  This is due to the fact that $g_{r r}$ is unchanged by the azimuthal fluid flow. This is a consequence of not having to satisfy the field equations, and is different from the Kerr case where $g_{r r}$ is affected by the rotation rate of the hole. 

For the Kerr type black hole, we have a rotational Killing vector $m=\partial /\partial \varphi$ as well as a time translational Killing vector $k$. The Killing vector that generates the horizon is \begin{equation}\ell=m+k= \frac{\partial}{\partial t} + \Omega_{H}  \frac{\partial}{\partial \phi} \end{equation} 
where $\Omega_{H} = 2\mathcal{W}/ \left( \sqrt{\alpha} \beta \right)$ is the coordinate angular velocity on the horizon. The surface gravity for the Kerr type black hole is the same as for the radial flow metric and is given by Eq.\ref{surfacegravity}.

 Although the surface gravity of the Kerr type hole is the same as the surface gravity of the radial flow hole they have different areas. The area of the Kerr type hole is 
\begin{equation}  \begin{split}A= &  \int \limits_{r=r_{H}}^{\null} \sqrt{g_{\theta \theta} g_{\varphi \varphi}} \, d\theta d\varphi  \\
& = \int \limits_{0}^{\pi}\frac{4\,\alpha \,{\beta }^3}  {{{\sqrt{ c^4\,{\beta }^2\,{\csc (\theta )}^2-4\,c^2\,{\mathcal{W} }^2\,
        \left( {\mathcal{W} }^2 - 
          \left(  \alpha -5  \right) \,{\beta }^2 \right) 
}}}} d\theta \label{areaintegral} \\
&= \frac{16\,\pi \,\alpha \,
    {\beta }^3\,
    \arctan (\frac{{\sqrt{  
           4\,\mathcal{W}^2\,
            \left( -5 + \alpha 
             \right) \,{\beta }^2-4\,\mathcal{W}^4}}}
        {c\,\beta })}{c\,
    {\sqrt{  
        4\,\mathcal{W}^2\,
         \left( -5 + \alpha  \right)
           \,{\beta }^2-4\,\mathcal{W}^4}}} \\
  \end{split}  , \end{equation}
if \begin{equation}\frac{{\mathcal{W }}^2\,
     \left( 5 - \alpha  + 
       \frac{{\mathcal{W} }^2}{{\beta }^2} \right) }{c^2}
   < \frac{1}{4} \label{bound} \end{equation}
where for simplicity we have set $\epsilon_{0}=0$.

The bound on $\mathcal{W}$ comes from the requirement that the denominator in Eq. \ref{areaintegral} be positive.  The bound on $\mathcal{W}$ is also the the condition need to ensure that $\sqrt{g_{\theta \theta} g_{\varphi \varphi}}$ is positive. We denote the value of $\mathcal{W}$ that saturates this bound by $\mathcal{W}_{c}.$

To calculate the effective mass of the Kerr type black hole we evaluate the Komar invariant integral \cite{Komar:1959,CohenAnddeFelice:1984}
\begin{equation}M =\frac{1}{8\pi G} \oint \limits_{\partial \Sigma}^{\null} \star d \xi_{t} \label{Kerrtypemass}, \end{equation}
where $\star d \xi_{t}$ is the Hodge dual of $ d \xi_{t} = g_{t i} dx^{i}$ and $\partial \Sigma$ is a space-like surface of constant $t$ and $r=r_{H}$. Evaluating Eq. \ref{Kerrtypemass} gives 
\begin{equation}
M= 2\frac{ \beta^{4}c^{2}}{ G} \int \limits_{0}^{\pi} \frac{\left(c^2 + \mathcal{W}^2 \alpha - \mathcal{W}^2 \alpha \cos (2 \theta) \right)
    \sin (\theta)}{\left| f(\mathcal{W}) \right|^{1/2} f(\mathcal{W}) }d \theta
\label{KerrHoleMass}
\end{equation} where 
\begin{equation} \begin{split} f(\mathcal{W})= & 
                  \left(c^2 + 2 \mathcal{W}^2
                        \left( \alpha-5 \right) \right)\, {\beta}^2 -2 \mathcal{W}^4 \\
 &+2 \mathcal{W}^2
                \left(\mathcal{W}^2 - \left(\alpha-5 \right) {\beta}^2 \right) \cos (2 \theta)   \\ \nonumber \end{split}.
\end{equation}

Unlike the radial flow case, the exterior mass given by Eq. \ref{exterior} does not go to zero even for an infinitely large container. This is due to the slow $1/r$ fall-off of the angular flow. 

The angular momentum of the hole is\cite{Komar:1959,CohenAnddeFelice:1984}
\begin{equation}J =-\frac{1}{16\pi G} \oint \limits_{\partial \Sigma}^{\null} \star d \xi_{\varphi} \label{angularmomentum}, \end{equation} where $d \xi_{\varphi}=g_{\varphi i} dx^{i}.$

Evaluating Eq.\ref{angularmomentum} gives 
\begin{equation}
J=-\frac{1}{16\pi G} \int \limits_{0}^{\pi} \frac{\left(8 \mathcal{W} \sqrt{\alpha} \beta^{3}\left(2\mathcal{W}^{2}-\left(\alpha-10 \right)\beta^{2} \right) \right)
    \sin (\theta)^{3}}{\left| f(\mathcal{W}) \right|^{1/2} f(\mathcal{W}) }d \theta
\label{angularmomentumintegral}\end{equation}

\begin{figure}
\includegraphics[scale=.8]{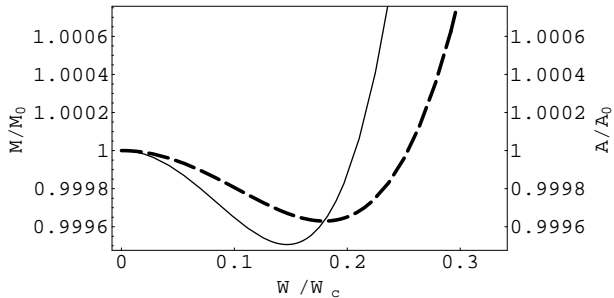}
\caption{Hole mass and area as a function of $\mathcal{W}$ with $\beta=1$ and $\alpha= 1\times 10^{7}$. The solid curve shows the hole mass plotted in reduced units showing $M/M_{0}$ as a function of $\mathcal{W}/\mathcal{W}_{c}$. The heavy dashed curve is $A/A_{0}$ where $M_{0}$and $A_{0}$ are the mass and area of the hole when $\mathcal{W}=0$. Notice that the curves intersect at  $\mathcal{W}/\mathcal{W}_{c} \approx 0.18 $ where  the angular momentum of the  hole has a non-trivial zero. }
\label{holemassfig}
\end{figure}

Although the integrals in  Eqs. \ref{KerrHoleMass} and \ref{angularmomentumintegral} can be done explicitly, the results are not particularly instructive.  However, from the integrand in \ref{angularmomentumintegral} we see that $J=0$ when $\mathcal{W}=\beta \sqrt{\left(\alpha-10\right)/2}$.
A plot of the hole mass  and area as a function of $\mathcal{W}$ is shown in Fig. \ref{holemassfig}.

From the definitions of the mass, angular momentum and surface gravity of the hole one can show that \cite{BardeenCarterHawking:1973}
\begin{equation}
M = \frac{\kappa A}{4 \pi G} +2\Omega_{H} J
\label{MassRelation} \end{equation}
\\

One can also verify the relation between the mass, angular momentum and area by explicit computation.

From Eq. \ref{MassRelation} and from the fact that $M=M(A,J)$ is a homogeneous function of degree 1/2 it follows \cite{Townsend:1997} that \begin{equation}
dM=\frac{\kappa}{8\pi G} dA+\Omega_{H} dJ
\end{equation} which is the first law of black hole thermodynamics.

In summary, we have shown that it is possible (at least in principle) to create an optical analog of a black hole which possesses all the important properties of a black hole, namely an event horizon which is a Killing horizon.
\bibliography{Optical_Black_Holes}

\begin{thebibliography}{14}
\expandafter\ifx\csname natexlab\endcsname\relax\def\natexlab#1{#1}\fi
\expandafter\ifx\csname bibnamefont\endcsname\relax
  \def\bibnamefont#1{#1}\fi
\expandafter\ifx\csname bibfnamefont\endcsname\relax
  \def\bibfnamefont#1{#1}\fi
\expandafter\ifx\csname citenamefont\endcsname\relax
  \def\citenamefont#1{#1}\fi
\expandafter\ifx\csname url\endcsname\relax
  \def\url#1{\texttt{#1}}\fi
\expandafter\ifx\csname urlprefix\endcsname\relax\def\urlprefix{URL }\fi
\providecommand{\bibinfo}[2]{#2}
\providecommand{\eprint}[2][]{\url{#2}}

\bibitem[{\citenamefont{Fresnel}(1818)}]{Fresnel:1818}
\bibinfo{author}{\bibfnamefont{A.~J.} \bibnamefont{Fresnel}},
  \bibinfo{journal}{Ann. Chim. Phys.} \textbf{\bibinfo{volume}{9}},
  \bibinfo{pages}{57} (\bibinfo{year}{1818}).

\bibitem[{\citenamefont{Lorentz}(1985)}]{Lorentz:1895}
\bibinfo{author}{\bibfnamefont{H.~A.} \bibnamefont{Lorentz}},
  \emph{\bibinfo{title}{Versuch einer Theorie der elektrischen und optischen
  Erscheinungen von bewegten K\"{o}rpern}} (\bibinfo{publisher}{Brill},
  \bibinfo{address}{Leiden}, \bibinfo{year}{1985}).

\bibitem[{\citenamefont{Gordon}(1923)}]{Gordon:1923}
\bibinfo{author}{\bibfnamefont{W.}~\bibnamefont{Gordon}},
  \bibinfo{journal}{Ann. Phys.} \textbf{\bibinfo{volume}{72}},
  \bibinfo{pages}{421} (\bibinfo{year}{1923}).

\bibitem[{\citenamefont{Hau et~al.}(1999)\citenamefont{Hau, Harris, Dutton, and
  Behroozi}}]{Hau:1999}
\bibinfo{author}{\bibfnamefont{L.~V.} \bibnamefont{Hau}},
  \bibinfo{author}{\bibfnamefont{S.}~\bibnamefont{Harris}},
  \bibinfo{author}{\bibfnamefont{Z.}~\bibnamefont{Dutton}}, \bibnamefont{and}
  \bibinfo{author}{\bibfnamefont{C.~H.} \bibnamefont{Behroozi}},
  \bibinfo{journal}{Nature} \textbf{\bibinfo{volume}{397}},
  \bibinfo{pages}{594} (\bibinfo{year}{1999}).

\bibitem[{\citenamefont{Leonhardt and Piwnicki}(1999)}]{LeonhardtPiwnicki:1999}
\bibinfo{author}{\bibfnamefont{U.}~\bibnamefont{Leonhardt}} \bibnamefont{and}
  \bibinfo{author}{\bibfnamefont{P.}~\bibnamefont{Piwnicki}},
  \bibinfo{journal}{Phys. Rev. A} \textbf{\bibinfo{volume}{60}},
  \bibinfo{pages}{4301} (\bibinfo{year}{1999}).

\bibitem[{\citenamefont{Leonhardt and Piwnicki}(2000)}]{LeonhardtPiwnicki:2000}
\bibinfo{author}{\bibfnamefont{U.}~\bibnamefont{Leonhardt}} \bibnamefont{and}
  \bibinfo{author}{\bibfnamefont{P.}~\bibnamefont{Piwnicki}},
  \bibinfo{journal}{Phys. Rev. Lett.} \textbf{\bibinfo{volume}{84}},
  \bibinfo{pages}{822} (\bibinfo{year}{2000}).

\bibitem[{\citenamefont{Hawking}(1972)}]{Hawking:1972}
\bibinfo{author}{\bibfnamefont{S.~W.} \bibnamefont{Hawking}}, in
  \emph{\bibinfo{booktitle}{Black Holes, Les Houches 1972}}, edited by
  \bibinfo{editor}{\bibfnamefont{C.}~\bibnamefont{DeWitt}} \bibnamefont{and}
  \bibinfo{editor}{\bibfnamefont{B.~S.} \bibnamefont{DeWitt}},
  \bibinfo{organization}{Summer School of Theoretical Physics of the University
  of Grenoble} (\bibinfo{publisher}{Gordon and Breach}, \bibinfo{address}{New
  York, N.Y.}, \bibinfo{year}{1972}), pp. \bibinfo{pages}{1--56}.

\bibitem[{\citenamefont{Felice and Clarke}(1990)}]{DeFelice&Clarke}
\bibinfo{author}{\bibfnamefont{F.~D.} \bibnamefont{Felice}} \bibnamefont{and}
  \bibinfo{author}{\bibfnamefont{C.~J.~S.} \bibnamefont{Clarke}},
  \emph{\bibinfo{title}{Relativity on Curved Manifolds}}
  (\bibinfo{publisher}{Cambridge University Press},
  \bibinfo{address}{Cambridge}, \bibinfo{year}{1990}).

\bibitem[{\citenamefont{Hawking}(1975)}]{Hawking:1975}
\bibinfo{author}{\bibfnamefont{S.~W.} \bibnamefont{Hawking}},
  \bibinfo{journal}{Comm. Math. Phys.} \textbf{\bibinfo{volume}{43}},
  \bibinfo{pages}{199} (\bibinfo{year}{1975}).

\bibitem[{\citenamefont{Joshi}(1993)}]{Joshi:1993}
\bibinfo{author}{\bibfnamefont{P.~S.} \bibnamefont{Joshi}},
  \emph{\bibinfo{title}{Global Aspects of Gravitation and Cosmology}}
  (\bibinfo{publisher}{Oxford University Press}, \bibinfo{address}{Oxford},
  \bibinfo{year}{1993}).

\bibitem[{\citenamefont{Bardeen et~al.}(1973)\citenamefont{Bardeen, Carter, and
  Hawking}}]{BardeenCarterHawking:1973}
\bibinfo{author}{\bibfnamefont{J.}~\bibnamefont{Bardeen}},
  \bibinfo{author}{\bibfnamefont{B.}~\bibnamefont{Carter}}, \bibnamefont{and}
  \bibinfo{author}{\bibfnamefont{S.}~\bibnamefont{Hawking}},
  \bibinfo{journal}{Comm. Math. Phys.} \textbf{\bibinfo{volume}{31}},
  \bibinfo{pages}{161} (\bibinfo{year}{1973}).

\bibitem[{\citenamefont{Komar}(1959)}]{Komar:1959}
\bibinfo{author}{\bibfnamefont{A.}~\bibnamefont{Komar}},
  \bibinfo{journal}{Phys. Rev.} \textbf{\bibinfo{volume}{113}},
  \bibinfo{pages}{1934} (\bibinfo{year}{1959}).

\bibitem[{\citenamefont{Cohen and de~Felice}(1984)}]{CohenAnddeFelice:1984}
\bibinfo{author}{\bibfnamefont{J.}~\bibnamefont{Cohen}} \bibnamefont{and}
  \bibinfo{author}{\bibfnamefont{F.}~\bibnamefont{de~Felice}},
  \bibinfo{journal}{J. Math. Phys.} \textbf{\bibinfo{volume}{25}},
  \bibinfo{pages}{992} (\bibinfo{year}{1984}).

\bibitem[{\citenamefont{Townsend}(1997)}]{Townsend:1997}
\bibinfo{author}{\bibfnamefont{P.~K.} \bibnamefont{Townsend}},
  \bibinfo{journal}{arXiv:gr-qc/9707012} p. \bibinfo{pages}{112}
  (\bibinfo{year}{1997}).

\end{thebibliography}

 \end{document}